\documentclass[smallextended]{svjour3}       
\smartqed
\usepackage{graphicx,epsfig}
\usepackage{amsmath}
\usepackage{amssymb}
\usepackage{epstopdf}
\usepackage{array}
\usepackage{float}

\newcommand{\be}{\begin{equation}}
\newcommand{\ee}{\end{equation}}
\newcommand{\bea}{\begin{eqnarray}}
\newcommand{\eea}{\end{eqnarray}}

\newcommand{\Rmnum}[1]{\expandafter\@slowromancap\romannumeral #1@}

\usepackage{color}

\begin{document}

\title{A quantum algorithm for solving systems of nonlinear algebraic equations}

\author{Peng Qian         \and
        Weicong Huang     \and
        Guilu Long
}


\institute{Peng Qian \and Weicong Huang
              \at
              State Key Laboratory of Low-dimensional Quantum Physics and Department of Physics, Tsinghua University, Beijing 100084, China\\
              Beijing National Research Center for Information Science and Technology, Beijing 100084, China\\
              Collaborative Innovation Center of Quantum Matter, Beijing 100084, China\\
              Beijing Academy of Quantum Information Sciences, Beijing 100193, China\\
              \email{qianp@itp.ac.cn,weiconghuang@mail.tsinghua.edu.cn}\\       
            Guilu Long \at
              State Key Laboratory of Low-dimensional Quantum Physics and Department of Physics, Tsinghua University, Beijing 100084, China\\
              Beijing National Research Center for Information Science and Technology, Beijing 100084, China\\
              School of Information and Technology, Tsinghua University, Beijing 100084, China\\
              \email{gllong@tsinghua.edu.cn}    
}


\date{Received:}

\maketitle

\begin{abstract}
We propose a quantum algorithm to solve systems of nonlinear algebraic equations. In the ideal case the complexity of the algorithm is linear in the number of variables $n$, which means our algorithm's complexity is less than $O(n^{3})$ of the corresponding classical ones when the set of equations has many variables and not so high orders. Different with most popular way of representing the results by state vector, we get the results in computation basis which is readable directly.
\end{abstract}

\section{Introduction}
\label{sect:Intro}
As one of the most important technology in the future of human beings, quantum computation has attracted more and more attention for its way to perform computation by harnessing quantum mechanics that classical computations can not. It is well known that in some problems quantum algorithm supply a remarkable speedup over their classical counterpart, for example, Shor's factoring algorithm\cite{Shor1999Polynomial}, Grover's searching algorithm\cite{Grover1996A,Long2001}, simulating various quantum systems\cite{Lloyd1996Universal,Gerritsma2009Quantum,Jordan2011Quantum}, HHL's algorithm for linear equations system\cite{Harrow2009Quantum} and so on.

In recent years, a number of researches concentrate on simulating Hamiltonian or solving linear system problems by quantum algorithm\cite{Lloyd2014Quantum,Berry2014Exponential,Berry2017Quantum}. Notice that these algorithms solving linear systems tend to represent results by the state vector evolved by quantum computer instead of computation basis used in early times. On one hand this way saves amounts of computing resource for saving lots of qubits. On the other hand it is hard to extract and read out numerical information. More than that, it's hard to realize nonlinear functions acted on these state vector since unitary operators transform the probability amplitudes linearly. Therefore for nonlinear problems, we still do not get appropriate algorithms yet.

In fact Shor in his famous and great algorithm used computation basis to represent results to solve the nonlinear problem --- order finding\cite{Shor1999Polynomial}. We learn that in this way it is easy to realize nonlinear function operation and extract results by just measuring computation basis and reading out digital numbers. Based on these, we develop a new methods to solve systems of nonlinear algebra equations. In essence it searches the whole space of computation basis to find approximation solutions to equations. However thanks to superposition principle it finds solution much faster than ever. It's kind of similar with Grover's algorithm while our method finds out multi-variables that satisfy multiple conditions.

Actually, our method is an application of duality quantum computation paradigm proposed first by Long\cite{Gui2005The,duallong} which realizes the non-unitary operator by combining several unitary gates through a number of ancilla qubits. The main idea comes from Double-slit interference. It imagines the state in quantum computation is a wave function in real space, passing through the multi-slits each of which corresponds to one unitary operator acting on the wave function, then intervening at some point behind multi-slits. Many algorithms can be realized through this paradigm, especially for non-unitary operator that can be decomposed into sum of unitary ones\cite{dualopen,Wei2017Realization,Wei2017quantumcom}. We use this paradigm to make solution found quickly and enhance the probability amplitude of the corresponding computation basis in total state vector. We also use some oracles based on the equations to be solved. In general, basic arithmetic operations have been realized in computation basis\cite{Vedral1996Quantum}. We treat them as feasible and reliable. What we concern is complexity of them and how they influence the total speed of the algorithm.

The rest of article is organized as follows: we first give the details of our algorithm. Then we analyze the complexity of our algorithm and compare it with the classical ones. In the end we will discuss possible extension of our algorithm and potential problems in practice before conclusion.

\section{Main algorithm}
\label{sect:main}
Assume we want to solve a group of algebraic equations:
\begin{equation}
\left\{f_{i}(x_{0},x_{1},\cdots,x_{n-1})=0\right\},i=0,1,2,\cdots,n-1
\end{equation}

First, we prepare n registers to represent every variable $x_{i}, i=0,1, \cdots, n-1$, that is, for initial state we set it as:
\begin{equation}
\left|\psi_{0}\right\rangle=\underbrace{\left|\overbrace{0\cdots0}^{N}\right \rangle\cdots\left|\overbrace{0\cdots0}^{N}\right \rangle}_{n}
\end{equation}
Here, we make this set according to what accuracy we need and the range for finding the solutions. If range is $(0,2^{m})$, then every register represents an integer if $N<m$ and a rational number if $N>m$ and latter $N-m$ digits represent decimal.

We apply Hadamard matrix $H^{\bigotimes Nn}$ to transform initial state to:
\begin{equation}
\left|\psi_{1}\right\rangle=H^{\bigotimes Nn}\left|\psi_{0}\right \rangle=\left(\sqrt{\frac{1}{2^{N}}}\right)^{n}\sum_{x_{0}\cdots x_{n-1}=0}^{2^{N}-1}\left|{x_{0}}\right \rangle\cdots\left|x_{n-1}\right\rangle
\end{equation}

We now add an ancilla register to make preparation for computation of oracle in the later computation:
\begin{equation}
\left|\psi_{1}\right\rangle\rightarrow\left(\sqrt{\frac{1}{2^{N}}}\right)^{n}\left(\sqrt{\frac{1}{N_{0}}}\right)\sum_{0}^{2^{N}-1}\left|{x_{0}}\right \rangle\cdots\left|x_{n-1}\right\rangle
\sum_{a=0}^{N_{0}-1}e^{i2\pi a/N_{0}}\left|a\right\rangle
\end{equation}
This register can be prepared by applying a quantum fourier transformation on a $\log{N_{0}}$ qubits state $\left|0\cdots 0\right\rangle$ which is easy to see.

For some oracle, we apply it on one state to get:
\begin{equation}
U_{f}\left|\Phi\right\rangle\rightarrow\sqrt{\frac{1}{x}}\sum^{x}\left|{x}\right \rangle\sqrt{\frac{1}{N_{0}}}\sum_{a=0}^{N_{0}-1}e^{i2\pi a/N_{0}}\left|a\oplus f(x)\right\rangle
\end{equation}
and redefine $a$ and get a extra common phase which reads:
\begin{equation}
\sqrt{\frac{1}{x}}\sum^{x}\left|{x}\right \rangle e^{i2\pi f(x)/N_{0}}\sqrt{\frac{1}{N_{0}}}\sum_{a=0}^{N_{0}-1}e^{i2\pi a/N_{0}}\left|a\right\rangle
\end{equation}
The procedure described above is so-called "phase kickback"\cite{Jordan2005Fast}, in which $N_{0}$ is chosen to be fit with our computation.

Now we put this aside for a moment and apply ${U_{f_{i}}}$ into the initial state for $i=0,\cdots,n-1$. Before doing that we first insert two zero registers into the state for storage of the computation results. Besides we add some control qubits for further application of oracles. Here, we take $N_{0}=2$:
\begin{equation}
\left|\psi_{1}\right\rangle\rightarrow\left(\sqrt{\frac{1}{2^{N}}}\right)^{n}\left(\sqrt{\frac{1}{2}}\right)^{n+1}\sum_{0}^{2^{N}-1}\left|{x_{0}}\right \rangle\cdots\left|x_{n-1}\right\rangle|0\rangle|0\rangle\sum_{a=0}^{1}e^{i\pi a}\left|a\right\rangle\sum_{\eta_{0}=0}^{1}\left|\eta_{0}\right\rangle\cdots\sum_{\eta_{n-1}=0}^{1}\left|\eta_{n-1}\right\rangle
\end{equation}

and then apply the unitary operator:
\begin{equation}
\label{eq:op}
\bigotimes_{i=0}^{n-1}
\{U_{f_{i}}^{-1}U_{\bar{f}_{i}}^{-1}(\sum_{j_{i}=0}^{1}M_{j_{i}}\left|j_{i}\right\rangle\left\langle j_{i}\right|)U_{\bar{f}_{i}}U_{f_{i}}\},
\end{equation}

where each $i$ corresponds a equation. For distinction we will write "I" as imaginary unit and "i" as index number.

Here, we take $U_{{f}_{i}}$ which makes computation of corresponding equation function $f_{i}$ and put the result into the first zero register, which reads $U_{f_{i}}|x\rangle|y\rangle\cdots|0\rangle=|x\rangle|y\rangle\cdots|f_{i}(x,y,\cdots)\rangle$. We take a oracle $\bar{f}_{i}$ we name with "check oracle" which counts whether the first few $\lambda$ digits of results are zero on range of solution and power of equation. If it is, the second zero register will not change. If not, the second zero register becomes $|1\rangle$. $M_{j_{i}}$ adds this result to ancilla register $\sum_{a=0}^{1}e^{I\pi a}\left|a\right\rangle$, changing it to $\sum_{a=0}^{1}e^{I\pi a}\left|(a\oplus j_{i}*\bar{f}_{i}\right)\rangle$. Then we use $U_{\bar{f}_{i}}$ and $U_{f_{i}}^{-1}$ to restore the zero register.

We use "phase kickback" and give the detailed process:

\bea
\begin{split}
&|\psi_{1}\rangle \xrightarrow{U_{f_{0}}}\sum_{0}^{2^{N}-1}\left|{x_{0}}\right \rangle\cdots\left|x_{n-1}\right\rangle|f_{0}(x_{0}\cdots x_{n-1})\rangle|0\rangle\sum_{a=0}^{1}e^{i\pi a}\left|a\right\rangle\sum_{\eta_{0}=0}^{1}\left|\eta_{0}\right\rangle\cdots\sum_{\eta_{n-1}=0}^{1}\left|\eta_{n-1}\right\rangle\\
&\xrightarrow{U_{\bar{f}_{0}}}\sum_{0}^{2^{N}-1}\left|{x_{0}}\right \rangle\cdots\left|x_{n-1}\right\rangle|f_{0}(x_{0}\cdots x_{n-1})\rangle|\bar{f}_{0}(x_{0}\cdots x_{n-1})\rangle\sum_{a=0}^{1}e^{i\pi a}\left|a\right\rangle\sum_{\eta_{0}=0}^{1}\left|\eta_{0}\right\rangle\cdots\sum_{\eta_{n-1}=0}^{1}\left|\eta_{n-1}\right\rangle\\
&\xrightarrow{\sum_{j_{0}=0}^{1}M_{j_{0}}\left|j_{0}\right\rangle\left\langle j_{0}\right|}\\
&\sum\left|{x_{0}}\right \rangle\cdots\left|x_{n-1}\right\rangle|f_{0}\rangle|\bar{f}_{0}\rangle(\sum_{a=0}^{1}e^{i\pi a}\left|a\oplus0*\bar{f}_{0}\right\rangle|0\rangle+\sum_{a=0}^{1}e^{i\pi a}\left|a\oplus1*\bar{f}_{0}\right\rangle|1\rangle)\sum_{\eta_{1}=0}^{1}\left|\eta_{1}\right\rangle\cdots\\
&\rightarrow\sum\left|{x_{0}}\right \rangle\cdots\left|x_{n-1}\right\rangle|f_{0}\rangle|\bar{f}_{0}\rangle\sum_{a=0}^{1}e^{i\pi a}\left|a\right\rangle(|0\rangle+e^{I\pi \bar{f}_{0}}|1\rangle)\sum_{\eta_{1}=0}^{1}\left|\eta_{1}\right\rangle\cdots\\
&\cdots\xrightarrow{U_{f_{0}}^{-1}U_{\bar{f}_{0}}^{-1}}\cdots\xrightarrow{U_{f_{1}}^{-1}U_{\bar{f}_{1}}^{-1}(\sum_{j_{1}=0}^{1}M_{j_{1}}\left|j_{1}\right\rangle\left\langle j_{1}\right|)U_{\bar{f}_{1}}U_{f_{1}}}\cdots\xrightarrow{U_{f_{1}}^{-1}U_{\bar{f}_{1}}^{-1}(\sum_{j_{n-1}=0}^{1}M_{j_{n-1}}\left|j_{n-1}\right\rangle\left\langle j_{n-1}\right|)U_{\bar{f}_{n-1}}U_{f_{n-1}}}\\
&\rightarrow\sum\left|{x_{0}}\right \rangle\cdots\left|x_{n-1}\right\rangle|0\rangle|0\rangle\sum_{a=0}^{1}e^{i\pi a}\left|a\right\rangle(|0\rangle+e^{I\pi \bar{f}_{0}}|1\rangle)\cdots(|0\rangle+e^{I\pi \bar{f}_{n-1}}|1\rangle)
\end{split}
\eea

We again apply a Hadamard operator $H^{\bigotimes n}$  on ancilla qubits $\left|j_{0}\right\rangle\cdots|j_{n-1}\rangle$. For those qubits satisfy $\bar{f}_{0}\cdots\bar{f}_{n-1}=0$, it looks like this:
\be
\sum_{0}^{2^{N}-1}|x_{0}^{*}\rangle \cdots |x_{n-1}^{*}\rangle|0\rangle|0\rangle\sum_{a=0}^{1}e^{I\pi a}\left|a\right\rangle|0\rangle.
\ee

\begin{figure*}
\includegraphics[width=0.75\linewidth]{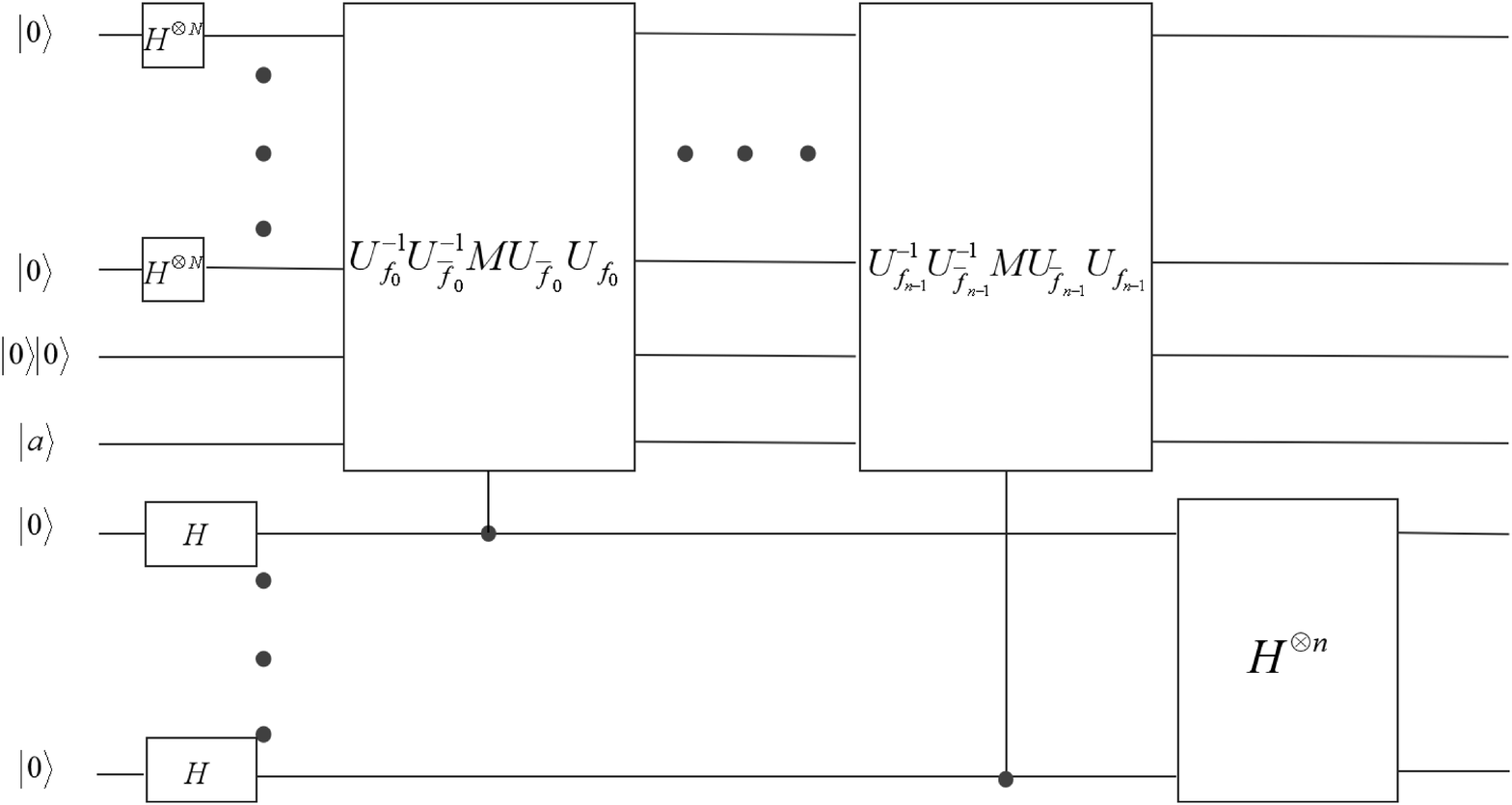}
\centering
\caption{circuit}
\label{figgg2}
\end{figure*}

While qubits presenting other variable values will get an zero amplitude in this ancilla $|0\rangle$ for orthogonality. This can be seen easily that $(|0\rangle+|1\rangle)\cdots(|0\rangle+|1\rangle)$(solutions' ancilla qubits) is orthogonal to others like $(|0\rangle+|1\rangle)\cdots(|0\rangle-|1\rangle\cdots$) and Hadamard rotation will not change this orthogonality so that qubits presenting other variable values will in ancilla qubits other than $|0\rangle$.
Actually, process above is a dual quantum computation in which we realize a complicated linear combination of unitary operators that cancel the undesired state and get the state we want.

Now we have to get into $\left|0\right\rangle$ in register so that only solution states are left behind. This can be done with amplitude amplification\cite{Brassard2000Quantum} or measurement(repeating until success). Obviously the best way is amplitude amplification.

We repeat acting $U_{0}=I-2I\bigotimes\left|0\right\rangle\langle0|$ and $U_{1}=I-2|\Psi\rangle\langle \Psi|$ where $|\Psi\rangle$ is the state we get in last step until we have a relatively large chance to measure $|0\rangle$ in ancilla qubits. If we first require the equations to be zero in first few $\lambda$ qubits then we loop these operations about $\sqrt{2^{\lambda}}$ and measure the ancilla qubits to make sure it is in $|0\rangle$,

However, we only get to approximation solution. We will then use gradient descent to find accuracy solutions. For simplicity we ignore all the normalized factors and add $l-(N-m)$ zero qubits to every variable register if our accuracy requirement is $2^{-l}$.
\begin{equation}
|x_{0}^{*}\rangle\underbrace{|0\cdots0\rangle}_{l-(N-m)}\cdots |x_{n-1}^{*}\rangle\underbrace{|0\cdots0\rangle}_{l-(N-m)}\sum_{N^{'}} e^{i\pi a/N^{'}}|a\rangle.
\end{equation}

We add every register $|\vec{x}_{i}^{*}\rangle$ to corresponding  $|y_{i}\rangle$, which is a complete set of corresponding accuracy, and rewrite variable register $|\vec{x}_{i}^{*}\rangle\underbrace{|0\cdots0\rangle}_{l+m-N}$ as $|\vec{x}_{i}^{*}\rangle$. We apply $U_{F}$ to get:
\begin{equation}
\sum_{y_{0}}|y_{0}+x_{0}^{*}\rangle\cdots\sum_{y_{n-1}}|y_{n-1}+x_{n-1}^{*}\rangle\sum_{N^{'}} e^{i\pi a/N^{'}}|a\rangle,
\end{equation}
where $F=\sum_{i}f_{i}^2$, $N^{'}$ is a parameter suited with accuracy\cite{Jordan2005Fast}.
We use "phase kickback" and do the quantum fourier transformation to get
\begin{equation}
|\frac{2^{m+l}}{s}\frac{\partial F}{\partial x_{0}}\rangle_{x_{0}^{*}}\cdots|\frac{2^{m+l}}{s}\frac{\partial F}{\partial x_{n-1}}\rangle_{x_{n-1}^{*}}
\end{equation}
where $s$ stands a range we set in oracle $U_{F}$ to bound derivatives of $F$.
We multiply this by a small negative constant $\alpha$ and add it to register $|x_{0}^{*}\rangle \cdots |x_{n-1}^{*}\rangle$ to get new value $|x_{0}^{**}\rangle \cdots |x_{n-1}^{**}\rangle$.
\begin{equation}
|x_{0}^{*}+\alpha\frac{\partial F}{\partial x_{0}}\rangle_{x_{0}^{*}}\cdots|x_{n-1}^{*}\frac{\partial F}{\partial x_{n-1}}\rangle_{x_{n-1}^{*}}
\end{equation}
We repeat the same procedure as above again and again until convergence. This is analogous to classical gradient descent. At last we just need measure register and read out the results.

\section{One simple example}
\label{sect:exa}
Here, we take a simple example to see how our method works. We take a ternary nonlinear systems of equations:
\bea
\begin{cases}
x^3 + y^2 - y + 2 z  = 35&\\
y^3 - x + 2zx = 50&\\
z^3 - z^2 + 2x - 2y = 20
\end{cases}
\eea
In real range $\{0,2^3\}$ with four decimal accuracy we have one solution $\{2.7689,3.2834,3.1370\}$. In our method, we first set accuracy to inter part which means we take $\lambda=3$ just need the inter part bits of equation to be zero. That is to say we take $\overline{f}$ to check the inter part of results of function oracle are whether all equal to zero. We set $N=6,m=3$. Then we take three equation function oracle to act on, for example, the state $|010.110\rangle|011.010\rangle|011.001\rangle|0\rangle|0\rangle(|0\rangle-|1\rangle)(|0\rangle+|1\rangle)^{\bigotimes 3}$(in decimal system$\{2.75,3.25,3.125\}$, an approximation solution). We found the qubit is unchanged after action. Here, we drop the normalization constant for simplicity. Then after Hadamard gate acted on last three ancilla qubits to get $|010.110\rangle|011.010\rangle|011.001\rangle|0\rangle|0\rangle(|0\rangle-|1\rangle)|0\rangle$. While for the other state, for example, $|011.010\rangle|010.010\rangle|011.100\rangle|0\rangle|0\rangle(|0\rangle-|1\rangle)(|0\rangle+|1\rangle)^{\bigotimes 3}$(in decimal system$\{3.25,2.25,3.125\}$)
\bea
\begin{split}
&|011.010\rangle|010.010\rangle|011.100\rangle|0\rangle|0\rangle(|0\rangle-|1\rangle)(|0\rangle+|1\rangle)^{\bigotimes 3}\\
&\xrightarrow{U_{f_{1}}}|011.010\rangle|010.010\rangle|011.100\rangle|10001.1011\rangle|0\rangle(|0\rangle-|1\rangle)(|0\rangle+|1\rangle)^{\bigotimes 3}\\
&\xrightarrow{U_{\bar{f}_{1}}}|011.010\rangle|010.010\rangle|011.100\rangle|10001.1011\rangle|1\rangle(|0\rangle-|1\rangle)(|0\rangle+|1\rangle)^{\bigotimes 3}
\end{split}
\eea
\bea
\begin{split}
&\xrightarrow{M_{1}|1\rangle\langle1|+M_{0}|0\rangle\langle0|}\\
&|011.010\rangle|010.010\rangle|011.100\rangle |10001.1011\rangle|1\rangle(|0\rangle-|1\rangle)|0\rangle(|0\rangle+|1\rangle)^{\bigotimes 2}\\
&+|011.010\rangle|010.010\rangle|011.100\rangle |10001.1011\rangle M_{1}|1\rangle(|0\rangle-|1\rangle)|1\rangle(|0\rangle+|1\rangle)^{\bigotimes 2}\\
\end{split}
\eea
$\bar{f}_{1}$ checks $|10001.1011\rangle$ to see if the inter part is zero. Obviously, $|10001.1011\rangle$ is not. Therefore the second zero register becomes $|1\rangle$. $M_{1}$ add this result to ancilla $(|0\rangle-|1\rangle)$, changing it to $-(|0\rangle-|1\rangle)$ while $M_{0}$ is equivalent to $I$:
\bea
\begin{split}
&\rightarrow|011.010\rangle|010.010\rangle|011.100\rangle |10001.1011\rangle|1\rangle(|0\rangle-|1\rangle)|0\rangle(|0\rangle+|1\rangle)^{\bigotimes 2}\\
&+|011.010\rangle|010.010\rangle|011.100\rangle|10001.1011\rangle|1\rangle(|1\rangle-|0\rangle)|1\rangle(|0\rangle+|1\rangle)^{\bigotimes 2}\\
&\rightarrow|011.010\rangle|010.010\rangle|011.100\rangle |10001.1011\rangle|1\rangle(|0\rangle-|1\rangle)(|0\rangle-|1\rangle)(|0\rangle+|1\rangle)^{\bigotimes 2}\\
&\xrightarrow{U_{\bar{f}_{1}}^{-1},U_{f_{1}}^{-1}}|011.010\rangle|010.010\rangle|011.100\rangle |0\rangle|0\rangle(|0\rangle-|1\rangle)(|0\rangle-|1\rangle)(|0\rangle+|1\rangle)^{\bigotimes 2}\\
&\cdots\\
&\rightarrow|011.010\rangle|010.010\rangle|011.100\rangle |0\rangle|0\rangle(|0\rangle-|1\rangle)(|0\rangle-|1\rangle)^{\bigotimes 3}
\end{split}
\eea
After Hadamard gate action, the state becomes $|011.010\rangle|010.010\rangle|011.100\rangle|0\rangle|0\rangle(|0\rangle-|1\rangle)|111\rangle$. Obviously, as we have proposed, all approximation solutions are in subspace $|0\rangle$. If it is not, then it will be in other subspace. From this we find we can quickly find approximation solution after amplitude amplification we describe above. Then we just need to run few iteration of gradient descent to get accurate solution.

\section{Complexity of the algorithm}
\label{sect:com}
For discussing the complexity of our algorithm, we need to count computing source each oracle costs. For basic arithmetic oracles like multiply we treat its complexity as $O(N^{2})$ \cite{Vedral1996Quantum} where $N$ is the size of our input. Note the highest order of group of equation is $h$ and the maximum number of terms of equations $t$, then the total number of basic operations we need for one equation oracle is $O(t h N^{2})$. The operator \eqref{eq:op} will need $O(n t h N^{2})$ elementary operations. Of course this analysis depends on how the elementary operation is defined and which oracle for $\bar{f}_{i}$ we choose. Here we just estimate the complexity order in ideal cases. After this we use amplitude amplification to get the approximate value. We need about $O(2^{\lambda/2})$ rotations. Therefore we need total $O(2^{\lambda/2}nt h N^{2})$ operations.

In gradient descent process, oracle $U_{F}$ costs $O(nth(l+m)^2)$.  We may set iteration times as a small number $c$ since we have got approximation solution.

At last we find that the total number of elementary operations in procedure will be $O(2^{\lambda/2}nt h N^{2}+c nth(l+m)^2)$. $\lambda$ is close to $hm$. Therefore complexity is linear in $n$ but exponential in $h$. For high order systems of equations we need to lower down approximation in first process to reduce complexity.

Now let's get a look at number of qubits our algorithm needs. For gradient descent process, we need $2n(l+m)$ qubits for registers. The ancilla qubits for oracle to make "phase kick-back" depend on oracle $U_{F}$. We note it as $\log{N^{'}}$. For accuracy it needs to be almost equal to $O(2h(m+l))$. Also for temporary store of results of $F$ we need $O(2h(m+l))$. The control qubits corresponding to the operator \eqref{eq:op} need about $O(n)$ qubits. What's more, there is a need of $O(hN)$ qubits for temporary store for computation results of function oracle of equations $f_{i}$. In sum, we need $O(4h(m+l)+hN)$ qubits for total store. The number depends heavily on the highest order of equations and accuracy requirement. Therefore for large order equations we may not obtain solution because computation result of function oracle is out of our computation source. However this situation is not so common for it means relatively large solution.

What about classical ones? The most common numerical method to solve these equations is Newton's iterative method which requires to solve the inverse matrix of Jacobian matrix for derivatives of each function with respect to each variable. This will cost about $O(n^{3})$ operators for $n$ variables. For further consideration, we still need to count complexity for each operator consisting of elementary arithmetic computation as for our quantum one. Then one will cost $O(ht (l+m)^{2})$ computations. In total it will have a complexity of $O(ht n^{3} (l+m)^{2})$ at one iteration. Considering that it will iterate many times until it converges to some point and this depends on choice of initial values sensitively, we think our algorithm will have a better performance when we have many number of variables.

In general our algorithm is similar to Grover's searching algorithm\cite{Grover1996A}. While in his algorithm it searches one solution to one condition. We think we extend to multi-variable problems. Notice that when the highest order is $1$, equations reduce to linear equations which have been researched frequently in recent year. In this case our algorithm will have a complexity of $O(n tN^2)$ better than classical ones, although it is not so well as HHL's\cite{Harrow2009Quantum}.

\begin{figure*}
\includegraphics[width=0.65\linewidth]{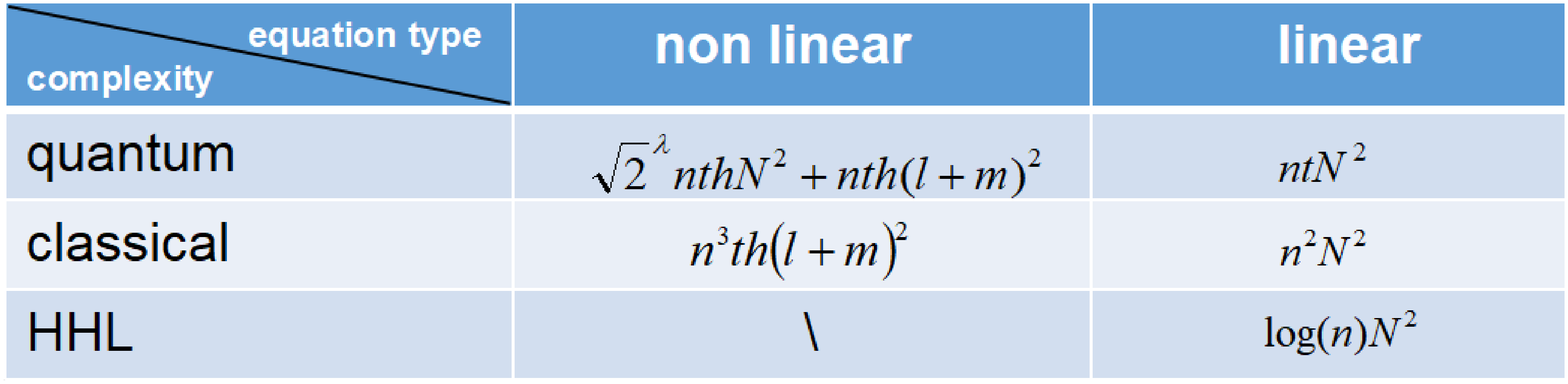}
\centering
\caption{comparison}
\label{figgg3}
\end{figure*}

We show a suitable way to solve nonlinear equations. However we know that there are many possible solution for one group of nonlinear equations and we may also meet the situation that we don't get the solution at all. In the former case, we have to repeat computation and measurement to get full solution. While in the latter case we may need to change the range for our initial register. That may cost more sources and we will hope that future's quantum computer may give us a enough powerful computation source.

\section{Conclusion}
\label{sect:conc}
In this paper we give a new algorithm to solve systems of nonlinear algebraic equations. Since we get a complexity linear in number of variable in systems of equations, we think our algorithm is better than classical ones. We can also change storage for results to adjust accuracy for results if we need as classical ones. Moreover, Based on computation basis we can read solution directly. One can also treat our algorithms as a multi-conditions and multi-variables search method. But for practice use we require future quantum computers to have enough qubits to store data in our algorithm. And for high order functions we're going to improve our algorithm and the related research is undertaken.
\begin{acknowledgements}
This work was supported by the National Basic Research Program of China under Grant Nos. 2017YFA0303700 and 2015CB921001, National Natural Science Foundation of China under Grant Nos. 61726801, 11474168 and 11474181.
\end{acknowledgements}




%
%

\end{document}